\documentclass[final,3p,times,twocolumn]{elsarticle}

\bibstyle{elsarticlenum-names}
 \usepackage{graphicx}
\usepackage{amssymb}
\usepackage{hyperref}

\journal{Journal of Magnetism and Magnetic Materials}

\begin{document}

%%\begin{frontmatter}


 %%\title{Title\tnoteref{label1}}

\title{Macrospin limit and configurational anisotropy in nanoscale Permalloy triangles }

 \author{L. Thevenard\corref{cor1}}
 \ead{l.thevenard@imperial.ac.uk}

\author{H. T. Zeng}
\author{D. Petit}
\author{R. P. Cowburn}

 \address{Blackett Laboratory, Physics Department, Imperial College London, Prince Consort Road, SW7 2BW London, United Kingdom}

%\address{}

\begin{abstract}

In Permalloy submicron triangles, configurational anisotropy - a higher-order form of shape anisotropy - yields three equivalent easy axes, imposed by the structures' symmetry order. Supported by micromagnetic simulations, an experimental method was devised to evaluate the nanostructure dimensions for which a Stoner-Wohlfarth type of reversal could be used to describe this particular magnetic anisotropy. In this regime, a straightforward procedure using an in-plane rotating field allowed us to quantify experimentally the six-fold anisotropy fields for triangles of different thicknesses and sizes.

\end{abstract}

%%\end{frontmatter}

 %%\linenumbers

\maketitle

\section{Introduction}
\label{sec:Introduction}

Much has been accomplished in previous years to shrink the feature size of nanomagnets in view of increasing the storage density of Magnetic Random Access Memories (MRAM) and hard disk drives, but also to develop materials exhibiting a magnetic anisotropy strong enough to resist thermal fluctuations at small dimensions, while remaining switchable by accessible magnetic fields \cite{Otte08,zimanyi08}. In the ubiquitous soft alloy Ni$_{81}$Fe$_{19}$ (Permalloy, Py), the magnetic anisotropy can, in certain cases, be tailored by the geometry of the nanostructure. For instance,  appropriately sized Py ellipses  have two stable states, where the magnetization lies along the longer in-plane dimension of the element. 

Another type of anisotropy exists in this material that remains to be harnessed to technological applications: configurational anisotropy  (CA) \cite{schabes88}. This phenomenon  is a direct effect of the rotational symmetry order ($n$) of a nanostruture on its magnetic anisotropy: triangles ($n$=3) evidence a six-fold anisotropy, squares ($n$=4)  a four-fold anisotropy, and pentagons ($n$=5) a ten-fold anisotropy\cite{cowburn00,jap09, vavassori}. Note that the necessity for easy and hard axes to present the same symmetry leads to frequency doubling for odd orders of $n$ \cite{cowburn98}. Configurational anisotropy relies on the fact that at small dimensions, a uniform magnetization cannot be sustained anymore in non-ellipsoidal structures, leading to a sizable deformation of the spin arrangement into an energetically more favorable state, balancing exchange and demagnetization energy costs. This effect could for instance allow the coding of multiple "bits" per nanostructure, or  be used for complex toggling or switching mechanisms in MRAM type structures.

To be of interest however, it is necessary to have an excellent control of this anisotropy. Different approaches have been proposed to measure higher-order magnetic anisotropies, such as Rotating field Magneto-Optic Kerr Effect (ROTMOKE) \cite{Mattheis99} in magnetic thin films, or Modulated Field Magneto-Optical Anisometry (MFMA) \cite{koltsov00,vavassori} in Supermalloy (Ni$_{81}$Fe$_{14}$Mo$_{5}$) and Fe nanostructures presenting CA. In both methods, large static fields are used to impose  macrospin-like behavior in the structure, in order to interpret the data within  a Stoner-Wohlfarth (SW) model and extract a value for the strength of this anisotropy. Here we present an alternative reliable and straight-forward experimental technique to obtain the six-fold anisotropy field of submicron triangles, and we use micromagnetic simulations to define a criterion estimating the maximum triangle dimensions up to which this parameter can indeed be defined.

\section{Samples }
\label{sec:sample}

Equilateral Py triangles were fabricated on Silicon by a 20kV electron beam lithography, followed by thermal evaporation and a lift-off process. A typical scanning electron micrograph of these triangles was presented in Ref. \cite{jap09}. Their widths and thicknesses varied between $e$=150-300~nm, and $t$=6-26~nm, and they were spaced by $e$ into large $25\times25\mu m^{2}$ arrays. In these nanostructures, configurational anisotropy is responsible for two possible magnetization states: a 'buckle' state where the global magnetization lies parallel to the edge, and the spins buckle from one corner to the other; or the 'Y-state', where $\vec{M}$ follows a triangle bisector, the spins splaying in (or out) from two corners toward the third \cite{koltsov00}. We have shown recently that for triangle dimensions above 100~nm wide/10~nm thick, only the buckle state can be stabilized \cite{jap09}, such that the $0^{\circ}$ [modulus $60^{\circ}$] directions are easy axes, and the $30^{\circ}$ [$60^{\circ}$] directions hard axes.

\section{Stoner-Wohlfarth switching astroid }
\label{sec:StonerWohlfarthCalculationOfH}

\begin{figure}[h]
	\centering
		\includegraphics[width=0.5\textwidth]{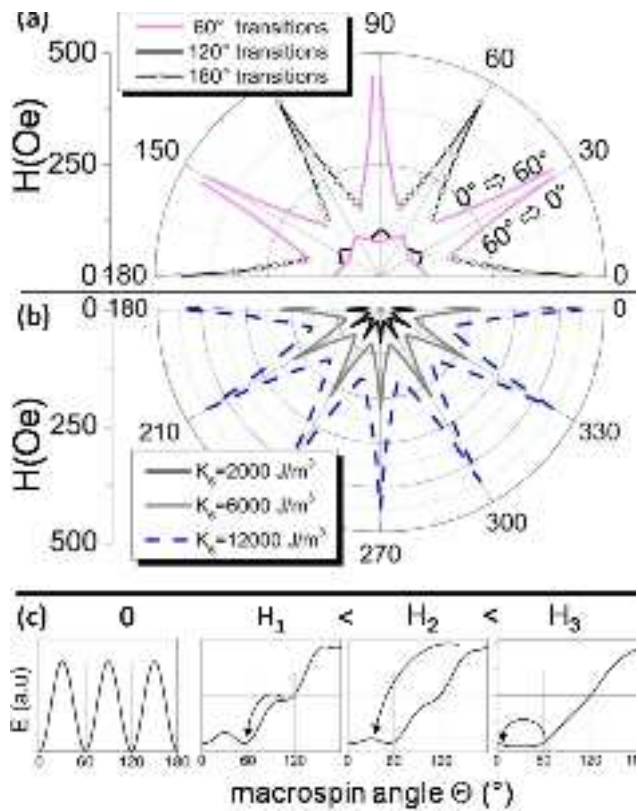}
	\caption{(a) Stoner-Wohlfarth switching astroid calculated for a six-fold anisotropy term  $K_{6}=12000~J/m^{3}$, with the  transitions $\pm60^{\circ}$, $\pm120^{\circ}$ or $180^{\circ}$. (b) Evolution of the envelope of the astroid for $K_{6}$  varying from 2000 to 12000$~J/m^{3}$. (c) Energy landscape for fields of increasing amplitude applied along $30^{\circ}$, leading to  $-60^{\circ}$ or $-120^{\circ}$ transitions. }
	\label{SW}
\end{figure}

While nanostructures exhibiting CA cannot, by definition, exhibit a macrospin behavior in the strict sense, it is tempting to compare their reversal to a Stoner-Wohlfarth (SW) model \cite{stoner,cowburn00}, where  the global magnetization is defined as $\vec{M}=(M_{s},\Theta)$, and the pseudo macrospin anisotropy is described by the simple energy functional $E_{6}=K_{6}sin^{2}(3\Theta)$, with $\Theta$ defined with respect to the base of the  triangle and $H_{6}=2K_{6}/M_{s}$. Here $K_{6}$ is positive, giving the buckles to be easy axes as is the case in our samples. To compute a SW switching astroid, a Zeeman term was then added to this expression and the total energy $E$ studied as a function of the applied field $\vec{H}=(H,\phi)$, with $\phi$ once again defined with respect to the base of the  triangle. The astroid  is constructed as follows, taking the example where the field $H$ is ramped along  the hard axis $\phi=30^{\circ}$ (Fig. \ref{SW}c). Starting from the $H=0$ energy functional $E(\Theta)$, a positive field is ramped up to $H_{1}$ for which a first transition is observed ($120^{\circ}$ to $60^{\circ}$) upon a local sign inversion of $d^{2}E/d\Theta^{2}$. The amplitude of this transition is $60^{\circ}$. With increasing field, the $180^{\circ}$ to $60^{\circ}$ transition occurs at $H_{2}$ (amplitude of $120^{\circ}$)  and finally $60^{\circ}$ to $0^{\circ}$ occurs at $H_{3}$ (once again a $60^{\circ}$ transition). For a field applied exactly along $30^{\circ}$, the $0^{\circ}$ and $60^{\circ}$ positions are  strictly speaking degenerate and the transition field then diverges; this actually leads to a computational artifact creating a slight asymmetry along the $30^{\circ}$ [modulus $60^{\circ}$] directions. The $60^{\circ}$ transition shown in Fig. \ref{SW}c (field $H_{3}$) will however be possible as soon as the field is brought away from the bisector. Applying the field in the opposite direction ($30^{\circ}$ + $180^{\circ}$) gives transitions  of equal amplitude at identical fields. For the field applied within $15^{\circ}$ of the easy  axes $0^{\circ}$ [$60^{\circ}$], an initial $60^{\circ}$  transition is followed at much higher field by a full $180^{\circ}$ transition. Note that there are two different types of $60^{\circ}$ transitions: the "lower-field" ones with maxima along $0^{\circ}$ [$60^{\circ}$], and the "higher-field ones", with maxima along $30^{\circ}$ [$60^{\circ}$]. For each field direction $\phi$, the transition fields $H_{i}$ are  evaluated in this way, and their locus plotted on the complete astroid, where transitions of different amplitudes $60$, $120$ or $180^{\circ}$ have been labeled (Fig. \ref{SW}a).

Very much like the SW astroid of an ellipse  presents maxima along its long and short axes directions, Fig. \ref{SW}a presents equal maxima along the easy and hard axes. The envelope of the full astroid is therefore a twelve-pointed star. The sole effect of increasing the anisotropy field is to expand the astroid, keeping all features identical, as shown in Fig. \ref{SW}b where three SW astroids were calculated for values of $K_{6}$ between 2000 and 12000~$J/m^{3}$ (for clarity, only the envelope of the astroid is shown).

\section{The rotating field method}
\label{sec:Experiment}

\subsection{Micromagnetic simulations}
\label{sec:MicromagneticSimulations}

Following these calculations, a possibility to extract a value of $K_{6}$ from experimental data would be to use an experimental switching astroid. The latter has been measured \cite{jap09}, but this method tends to yield fairly large error bars on the transition fields, as it requires to isolate specific transitions as small kinks in the hysteresis loops. Another solution consists of \textit{forcing} $60^{\circ}$  rotations of the magnetization in the triangles, for instance by using a large rotating field. By measuring the field amplitude required to observe a magnetization flip from one buckle to the next, the $60^{\circ}$ transition fields can be obtained and compared to the equivalent feature in the SW model.

\begin{figure}
	\centering
	\includegraphics[width=0.5\textwidth]{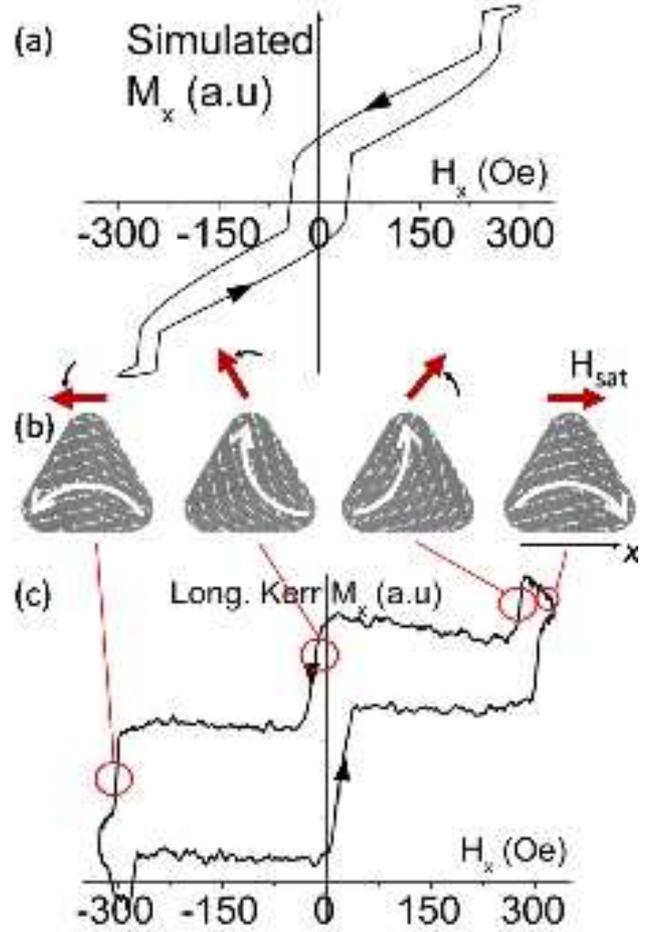}
	\caption{(a) and (b) [color on-line] Numerical simulations of the effect of a counterclockwise rotating field of constant amplitude on a 310nm wide triangle.  The projection along $x$ of the magnetization is plotted with the corresponding field direction (thick red arrow) and global magnetization orientation (white arrow).  (c) Experimental longitudinal Kerr magnetization plotted against the $x$ component of a counterclockwise rotating field $H_{rot}$=~324 Oe (sweeping direction given by the black arrows), for an array of 300~nm wide, 9~nm thick triangles.}
	\label{Rot}
\end{figure}

In order to test the validity of this approach, numerical simulations  (\textsc{OOMMF} package \cite{oommf}) were first done on 310~nm wide, 10~nm thick triangles, using a 5~nm cell size, $M_{s}=~800~kA/m$, $A=13~pJ/m$ and $\alpha=0.5$. A rounding of $\oslash=60$~nm was moreover included to take into account the physical rounding of our structures \cite{jap09}, and the triangular mesh was titled by $15^{\circ}$ to spread pixelation effects. A counterclockwise field of amplitude $H_{rot}$ was then applied, starting from the positive $x$ direction with the magnetization initialized in a $0^{\circ}$ buckle. Unsurprisingly, for low amplitude rotating fields ($H_{rot}\leq150$~Oe), no rotation of the global magnetization $\vec{M}$ was observed at all, whereas for $H_{rot}\geq1000$~Oe, $\vec{M}$ followed exactly the field. In the intermediate regime, the temporal evolution of the magnetization and the corresponding spin configuration are as shown in Fig. \ref{Rot}a,b (for clarity, the triangles were not shown tilted by $15^{\circ}$). Six abrupt $60^{\circ}$ rotations of $\vec{M}$ are observed during a full field cycle, with the magnetization flipping from one buckle to the next. The amplitude of the transitions in Fig. \ref{Rot}a  corresponds to the projection of the global magnetization drawn schematically by large white arrows in Fig. \ref{Rot}b. The simulations therefore seem to show that it is possible to address a particular transition, the $60^{\circ}$ rotation, by using a rotating field of carefully chosen amplitude.

\begin{figure}
	\centering
		\includegraphics[width=0.5\textwidth]{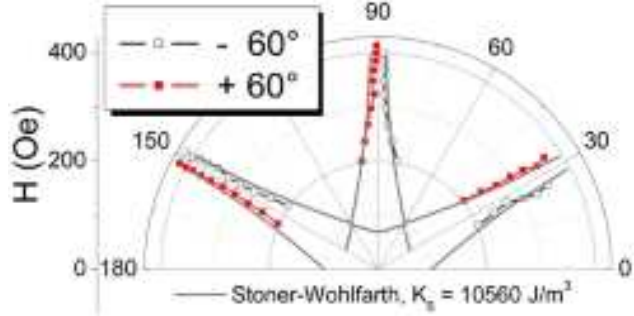}
	\caption{[color on-line] Polar plot of the rotating field amplitude versus the angle at which occurs a $60^{\circ}$ rotation of the global magnetization in 300~nm wide, 9~nm thick triangles. Clockwise (resp. counterclockwise) fields yield the $-60^{\circ}$ transitions, empty symbols (resp. $+60^{\circ}$, closed symbols). The solid line is a fit to the Stoner-Wohlfarth calculation of the $0-60^{\circ}$ transition, with $K_{6}=10560~J/m^{3}$.}
	\label{polar}
\end{figure}

\subsection{Experimental methods}
\label{sec:exp}

The effect of a 1 Hz rotating field  was then studied experimentally using Magneto-Optical Kerr Effect (MOKE) in the longitudinal geometry ($x$ axis in Fig. \ref{Rot}b). The laser was focused into a $5~\mu m$ spot on each array, and the field created by a combination of in-plane fields ($H_{x}$,$H_{y}$) from a quadrupole electromagnet. A typical hysteresis loop is presented in Fig. \ref{Rot}c, where the longitudinal magnetization $M_{x}$ is plotted against the $x$ component of a counterclockwise rotating field $H_{rot}=324$~Oe. The loop exhibits 6 very clear and sudden changes in $M_{x}$, in two sets of 3 symmetrical transitions (longitudinal field sweeping down or up). Comparing with the simulations, these can be identified as successive $60^{\circ}$ rotations of the magnetization occurring at specific projections along $x$ of the rotating field:  $h_{x1}=\pm270$~Oe, $h_{x2}=\pm17$~Oe and $h_{x3}=\pm291$~Oe. Because the amplitude of the field remains constant during a full cycle, it is immediate to calculate the angle of the field at the moment of the transition, using $\phi_{i}=cos^{-1}(h_{xi}/H_{rot})$, or $\phi_{i}=cos^{-1}(h_{yi}/h_{xi})$. This yields for the loop shown: $\phi_{1}=34^{\circ}~[180^{\circ}]$, $\phi_{2}=93^{\circ}~[180^{\circ}]$ and $\phi_{3}=154^{\circ}~[180^{\circ}]$. For a triangle in a $0^{\circ}$ buckle, a rotating field of 324~Oe will therefore induce a rotation toward the $60^{\circ}$ buckle just as it crosses the $\phi_{1}=34^{\circ}$ direction. Faraday effects and other perturbations tend to introduce an arbitrary slope to the loops, which we have subtracted for clarity;  the amplitude of the transitions are therefore not directly comparable with the simulations . 

\subsection{Results}
\label{sec:results}

For each array, $M(H)$ loops were measured under increasing values of $H_{rot}$, starting from the lowest one giving a 6-stepped loop. The data was then represented as a polar plot representing the rotating field amplitude as a function of the transition angles $\phi_{i}$ (Fig. \ref{polar}). Note that both $+60^{\circ}$ and $-60^{\circ}$ transitions are represented. They were obtained by applying the field counterclockwise or clockwise. The experimental $H_{rot}(\phi)$ data evidences the same characteristics as the computed higher-field  $60^{\circ}$ transitions either side of  $\phi=30^{\circ}$ [$60^{\circ}$] in the SW astroid. Mainly, if the magnetization lies along the bottom edge of the triangle ($0^{\circ}$ buckle), it can only flip by $\pm 60^{\circ}$ to the following buckle under the torque of a field applied along $\pm 30^{\circ}$ at the least. When this angle is increased, the field required to obtain a transition strongly decreases, and ceases to be measurable at about $\phi=\pm45^{\circ}$. This is naturally reminiscent of the magnetization reversal mechanism in ellipses, where the additional torque from an $H_{y}$ field perpendicular to $H_{x}$ greatly reduces the switching field.  

We then attempted to fit the experimental $H_{rot}(\phi)$ data to the higher-field $60^{\circ}$ transitions calculated in the SW astroid (Fig. \ref{SW}), the only adjustable parameter being  $K_{6}$. For the plot shown in Fig. \ref{polar} for instance (300~nm wide, 9~nm thick triangles), the best fit is thus obtained for $K_{6}=10560\pm300~J/m^{3}$, or expressed in term of anisotropy field, $H_{6}=264\pm8~Oe$ ($M_{s}=800~kA/m$). 

\section{Establishing a macrospin limit}
\label{sec:EstablishingAMacroSpinLimit}

This procedure was applied to triangles of different widths ($e=$150-300~nm), and thicknesses ($t=$6-26~nm). For all of these arrays, 6-steps hysteresis loops were easily obtained under a rotating field, and a $H_{rot}(\phi)$ polar plot of $60^{\circ}$ transitions readily extracted.  Before retrieving $H_{6}$ values from these data however, the validity limits of this method needs to be questioned. Indeed, while the 6-stepped hysteresis loops seem to warrant a macrospin-like behavior, and therefore legitimize the approach described above, it can be argued that the existence of a $60^{\circ}$ transition is not a good criteria for a macrospin behavior in our nanostrutures, being an easy, small amplitude 'sliding' of spins from one buckle to the next. A full $180^{\circ}$ reversal, or any procedure allowing the magnetization to relax would be a more reliable control test. The experimental switching astroid now becomes a useful tool, as it is measured by applying a saturating field along a direction $-\phi$, letting $\vec{M}$ relax while reversing the field direction, and observing the configurational changes of the spins leading to a full reversal of the magnetization along $+\phi$.

\begin{figure}
	\centering
	\includegraphics*[width=0.5\textwidth]{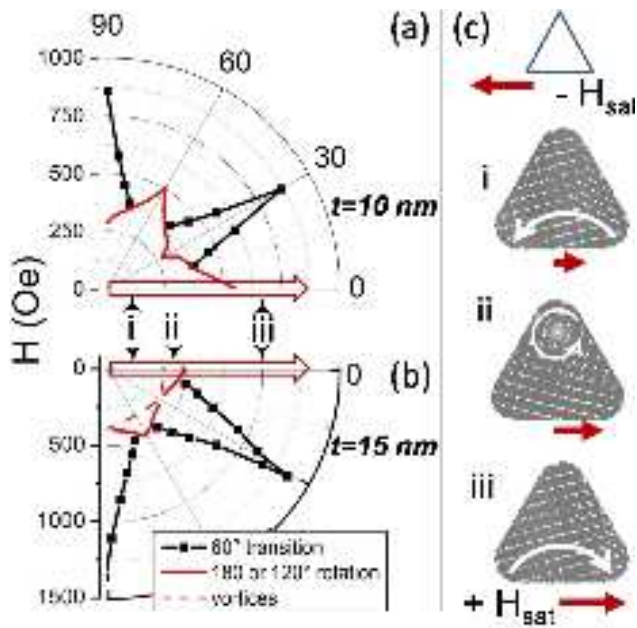}
	\caption{[color on-line] Micromagnetic simulation of  the switching astroids. The field $H$ is saturated along $-\phi$, and ramped back up along $+\phi$ (large empty red arrow), with the micromagnetic configuration evolving from (i) to (iii). Configuration (ii) is only observed for  $t$=15~nm. The switching fields are plotted versus $\phi$.  (a) 300~nm wide, 10~nm thick triangle: abrupt switching of $M$ for $H$ reversed along $\phi=0^{\circ}$. (b)  200nm wide, 15nm thick triangle: a new, low-field transition appears when increasing the thickness. (c) Spin configurations corresponding to the $\phi=0^{\circ}$ reversal for (a) and (b), with the global magnetization direction given by the white arrow, and the field direction by the full red arrow. }
	\label{oommf}
\end{figure}

A micromagnetic simulation of the  switching astroid was then done following the field sequence described above, and the  fields inducing abrupt changes of the magnetization plotted versus the direction of the applied field. The dimensions (thickness, width) investigated were identical to the experimental ones. For thicknesses $t\leq10$~nm at all triangle widths, as well as the $t$=15~nm, $e$=150~nm structure, a unique shape was obtained for the astroid, an example of which is shown in Fig. \ref{oommf}a (first quadrant of the simulation for the $t$=10~nm, $e$=300~nm triangle). Having been described in detail in Ref. \cite{jap09}, only the main differences and similarities with a macrospin type of reversal (Fig. \ref{SW}a) will be highlighted. For fields reversed along directions close to the easy axes $0^{\circ}$ [$60^{\circ}$], the magnetization does a complete $180^{\circ}$ reversal, going \textit{directly} from state (i) to (iii) in Fig. \ref{oommf}c. For fields close to the hard axes $30^{\circ}$ [$60^{\circ}$], it reverses in two steps, a $120^{\circ}$ transition  followed by a $60^{\circ}$ transition (closed symbols in Fig. \ref{oommf}a). The latter is the exact same transition observed under a rotating field, or calculated using the SW model. Contrary to the SW astroid, the peaks along $0^{\circ}$ [$60^{\circ}$] are lower than the ones at $30^{\circ}$ [$60^{\circ}$]. This is a direct consequence of  the complex and different micromagnetic  configurations in the easy and hard axes states. In both easy and hard-axis types of reversal, note that there is one less transition compared  to the SW model: for all field directions, the first $60^{\circ}$ transition does not occur in the micromagnetic simulations. Moreover, the locus of the $120^{\circ}$ transition fields has a quite different shape in both cases. On the one hand, the macrospin model identifies local energy minima, without favoring any initial or final configuration: the first $120^{\circ}$ to $60^{\circ}$ transition $H_{1}$ in Fig. \ref{SW}c can only occur if the initial configuration of $\vec{M}$ is along $120^{\circ}$ for instance. In the micromagnetic simulations on the other hand, the triangle is first fully saturated in one direction, thereby fixing the initial magnetic configuration, before an opposite field is applied.

When the triangle width $e$ is increased above 150~nm ($e\geq$200~nm) for $t\geq15$~nm, the envelope of the astroid remains the same, a 12-pointed star, but a new transition appears at low fields, as shown in Fig. \ref{oommf}b by the dashed line. Indeed, the simulations evidence for any applied field direction a passage by an intermediary vortex state (ii) shown in Fig. \ref{oommf}c. This vortex is eventually expelled as a  $60^{\circ}$ transition does occur at higher fields. Obtaining a 6-stepped hysteresis loop under rotating field  (series of $60^{\circ}$ transitions) is therefore not a good 'macrospin' criterion. Following the simulations however, a simple remanence loop along the $0^{\circ}$ direction of the triangle should  evidence the passage by a vortex state, seen as the presence or absence of a second step in the loop, and  determine whether it is legitimate to extract an $H_{6}$ value from the data obtained under a rotating field. 

\begin{figure}
	\centering
		\includegraphics[width=0.5\textwidth]{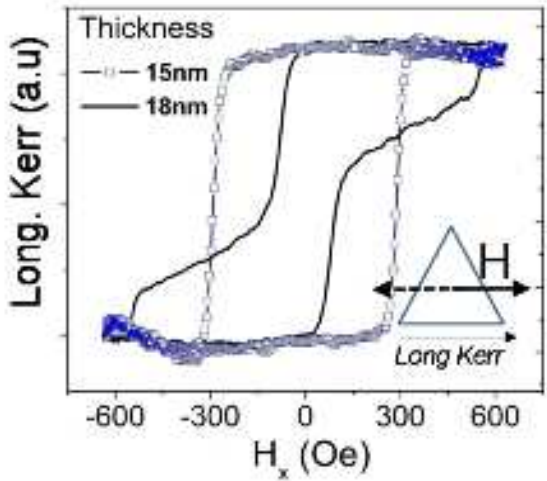}
	\caption{[color on-line] Longitudinal Kerr remanence loops for the field applied along the base of the triangles:  15~nm thick, 300~nm wide (open symbols), or 18~nm thick, 150~nm wide (full line). The double-step in the latter evidences the passage through a vortex state, and therefore the boundary of the macrospin behavior.}
	\label{loops}
\end{figure}

Remanence loops along the $0^{\circ}$ direction were measured  for the different arrays. A clear cut-off dimension appears below which the reversal is a single step $180^{\circ}$ reversal (thickness 15~nm, width 300~nm), as shown in Fig. \ref{loops}, open symbols. Above this value  (thickness 18~nm, width 150~nm, full line in Fig. \ref{loops}), a double step transition evidences the formation of a vortex, and therefore the boundary of applicability of our SW-based analysis of the six-fold anisotropy field. The cut-off dimension between macrospin and non-macrospin-like behaviors are therefore quite similar between simulations and experiments in the explored size range: 15nm thick, 200~nm wide for the first, 18~nm thick, 150~nm for the second. It may be argued that the zero-temperature nature of these computations lets the system enter shallow energy wells, such as the vortex configuration, whereas thermal fluctuations in the experiment leave them unnoticed. Within the explored dimensions, the volume of the nanostructure at the boundary can however be considered identical between simulations and experiment.

\section{Discussion and  Conclusions}
\label{sec:Discussion}

Following this measurement, six-fold anisotropy fields were   extracted from the MOKE data for triangles $e$=150-300~nm and $t$=6-15~nm, as summarized in Table \ref{tab:H6}. $H_{6}$ increases with thickness, ranging from 128~Oe to 706~Oe.  Where they can be compared, the values agree well with the anisotropy fields obtained in thinner triangles using MFMA \cite{cowburn00}; the thinnest sample presented in the present study for instance (6~nm thick, 300~nm wide) has an anisotropy field of $H_{6}=128\pm10~Oe$, very comparable to the 107~Oe obtained by MFMA in 5~nm thick, 270~nm wide triangles. The method presented here however, gives an estimation of the limit of a macrospin description of configurational anisotropy. Finally, these anisotropy fields are comparable to experimental uniaxial anisotropy fields $H_{un}$ obtained in Permalloy ellipses \cite{cowburn7067}, with the major difference that they are obtained in non-elongated structures via configurational anisotropy. 

{\footnotesize
	\begin{table}[!h]
	\centering
		\begin{tabular}{ c  c  c }
					 Thickness (nm)     & Width (nm)						&  $H_{6}$ (Oe) \\
	\hline	\hline 6            & 300                & 128 \\
		\hline                    & 150                & 161 \\
		 			 9                  & 200                & 317 \\
		 		                      & 300                & 264 \\
		\hline                    & 150                & 561 \\
		 			 15                 & 200                & 706 \\
					                    & 300                & 670 \\
		\end{tabular}
	\caption{Six-fold anisotropy fields $H_{6}$ extracted from rotating field loops, using $H_{6}=2K_{6}/M_{s}$.}
		\label{tab:H6}
\end{table}
}

  In elliptical or rectangular platelets (area $a*b$, thickness $t$), $H_{un}$ does not depend on exchange energy in first approximation, but mainly on the demagnetizing field $H_{demag}$ of the structure. Increasing $t$ creates more charges at the structure edges, and decreasing the width (keeping $a/b$ constant) brings these poles closer together, both leading to a larger $H_{demag}$ \cite{aharoni98,osborn45}. A monotonous increase of $H_{un}$ as a function of $t/a$ is then expected.  This is in part what is observed in the triangles: at fixed triangle width, the anisotropy field increases with $t$ (Table \ref{tab:H6}). At fixed triangle thickness however, $H_{6}$ first increases when the width decreases, and peaks at around 200~nm before decreasing. Interestingly, this latter trend with width is not observed for anisotropy fields  extracted from zero-temperature micromagnetic simulations (Fig. \ref{Rot}) following the procedure described in section \ref{sec:results}: the $H_{6}$ fields are in this case found to increase monotonously with increasing thickness and decreasing structure size. For the smaller structures ($e. g$ 9~nm thick, 150nm~wide) the anisotropy energy  $U=2M_{S}VH_{6}/n^{2}$ deduced from the experimental anisotropy fields  amounts to about 10$k_{B}T$ (volume $V$ of the triangle, symmetry order $n$ of the structure \cite{cowburn00}). Thermal fluctuations then become sufficient to overcome the anisotropy barrier, and $H_{6}$ starts to decrease as the structure is brought closer to the superparamagnetic regime. Note that configurational anisotropy is weak enough to let us observe the ferromagnetism-superparamagnetism transition at larger dimensions than in Permalloy ellipses \cite{cowburn03}.

	In conclusion, we have devised a simple experimental method to extract the six-fold anisotropy fields of Permalloy triangles presenting configurational anisotropy. Using a large rotating field to impose a particular $60^{\circ}$ transition, we created  conditions to compare our data to a Stoner-Wohlfarth model. Moreover, using micromagnetic simulations, we proposed a straightforward test to assess the validity of this approach for triangles of any given dimensions. Finally, we highlighted the specificity of configurational anisotropy in which the usual thickness over width dependence of the anisotropy field is not fully observed due to the proximity of the superparamagnetic regime. Equipped with this fundamental parameter characterizing the magnetic anisotropy, further work will focus on assessing the stray field interaction between these structures, in order to compare them to typical values of nanomagnets currently used in data storage technologies.



\end{document}